# Neutrino in Astrophysics and Cosmology

Zuxiang Dai

**Tutor:** Igor Tkachev

**Professor:** Christoph Schmid

| Proseminar in Theoretischer Physik | **ETH** *Zürich* |
|---|---|
| Kosmologie und Teilchenphysik | Date: 12. Jan. 2000 |

**Abstract.** At first I introduce the Neutrino in the standard Model, Then the Dirac and Majorana Masses. After introduce the See-Saw Mechanism, I discuss the neutrino oscillations and neutrino in astrophysics and cosmology. I finished this paper with a brief summary of the neutrino experiments.

**1     The standard model and the neutrino**

In 1961 Glashow proposed the SU(2)×U(1) group as a possible local symmetry of weak interactions. Then in year the 1967/1968 Weinberg and Salam proposed the spontaneously broken $SU(2)_L \times U(1)_Y$ gauge theory. Based on these theories, we identify the electromagnetic interaction and the weak interaction as two aspects of one interaction. After 't Hooft proved, in 1971, that the theory is renormalizable in less than or equal to four dimensions, and this year (1999) he is the Nobel prize winner for his work. Gross, Wilczek and Politzer proposed the SU(3) group for strong interaction in 1973. Since neutrino does not exhibit strong interactions, we will only concentrate on the $SU(2)_L \times U(1)_Y$ Model.

We know that, left-handed neutrinos are part of SU(2) doublets and right-handed neutrinos are SU(2) singlets.

$$L_i = \begin{pmatrix} \nu_{\ell_i} \\ \ell_i \end{pmatrix}, \quad \ell_i = \{e, \mu, \tau\} \tag{1.1}$$

From Gell-Mann and Nishijima we have the relation:

$$Q = T_3 + \frac{1}{2}Y \tag{1.2}$$

Q is the electromagnetic charge, Y is the U(1) hypercharge, and $T_3$ is the 3rd component of the weak isospin. We have a small table:

|   |   |   | T | $T_3$ | Y | Q |
|---|---|---|---|---|---|---|
| $\begin{pmatrix} \nu_e \\ e \end{pmatrix}_L$ | $\begin{pmatrix} \nu_\mu \\ \mu \end{pmatrix}_L$ | $\begin{pmatrix} \nu_\tau \\ \tau \end{pmatrix}_L$ | ½ | +½ | -1 | 0 |
|   |   |   |   | -½ |   | -1 |
| $e_R$ | $\mu_R$ | $\tau_R$ | 0 | 0 | -2 | -1 |

One notes that the right-handed neutrinos carry no SU(2)×U(1) quantum numbers. Experimentally we see that only purely left-handed neutrinos are produced by the weak interactions. Therefore, for the purpose of this paper, we will in most case ignore the presence of right-handed neutrinos.

Now we are interested in the determination of the number of neutrino species *i*. Here we study the Z-boson decay. We show at first the Breit-Wigner formula:

$$\sigma_{i \to f}(s) = 12\pi(\hbar c)^2 \cdot \frac{\Gamma_i \cdot \Gamma_f}{(s - M_Z^2 c^4)^2 + M_Z^2 c^4 \Gamma_{tot}^2}. \tag{1.3}$$

| | |
|---|---|
| $\sigma$… | cross section |
| $\Gamma_i$… | Decay width in input channel |
| $\Gamma_f$… | Decay width in output channel |
| $\sqrt{s}$ … | Center of mass energy |
| $M_Z$… | Mass of Z-boson |

And:

$$\Gamma_{tot}(Z^0) = \sum_{all\ fermions} \Gamma(Z^0 \to \bar{f}f). \tag{1.4}$$

When we use the nature unit with $\hbar = c = 1$ and $s = M_Z^2 c^4$ we have the peak cross-section $\sigma_o$:

$$\sigma_0 = 12\pi \cdot \frac{\Gamma_i \cdot \Gamma_f}{M_Z^2 \Gamma_{tot}^2}. \tag{1.5}$$

for the $e^+e^- \to$ hadrons we have:

$$\sigma_0 = 12\pi \cdot \frac{\Gamma_{lept} \cdot \Gamma_{had}}{M_Z^2 \Gamma_{tot}^2}. \tag{1.6}$$

Here we suppose all lepton channels have the same decay width, which is proved by the experiment on LEP and SLC, [6]:

$$\begin{aligned}
Z^o \to \quad & e^+ + e^- & & 3.366 \pm 0.008\ \% \\
& \mu^+ + \mu^- & & 3.367 \pm 0.013\ \% \\
& \tau^+ + \tau^- & & 3.360 \pm 0.015\ \% \\
& \nu + \bar{\nu} & & 20.01 \pm 0.16\ \% \\
& \text{hadrons} & & 69.90 \pm 0.15\ \%
\end{aligned} \tag{1.7}$$

We call it Lepton universality. Further more we have a formula for the hadron cross-section from [1]:

$$\sigma_{had} = \sigma_0 BW(s)(1 - \delta_{QED}(s)). \tag{1.8}$$

Where the $\delta_{QED}(s)$ is a computable initial state bremsstrahlung correction from QED, while the Breit-Wigner factor is given by:

$$BW(s) = \frac{s \cdot \Gamma_{tot}^2}{(s - M_Z^2)^2 + s^2 \Gamma_{tot}^2 / M_Z^2}. \tag{1.9}$$

And we get from [11]:

$$\sigma_o = (41.491 \pm 0.058)\ nb \tag{1.10}$$

We know 1 barn = 1 b = $10^{-28}$ m$^2$.

From [11] we see the ratios:

$$R_e := \Gamma_{had}/\Gamma_{ee},\ R_\mu := \Gamma_{had}/\Gamma_{\mu\mu}\ \text{and}\ R_\tau := \Gamma_{had}/\Gamma_{\tau\tau}. \tag{1.11}$$

Here $\Gamma_{ee}$, $\Gamma_{\mu\mu}$ and $\Gamma_{\tau\tau}$ are the partial widths of the Z-boson for the decays $Z \to e^+ + e^-$, $Z \to \mu^+ + \mu^-$ and $Z \to \tau^+ + \tau^-$. Due to the large mass of the $\tau$ lepton, a small difference of 0.2 % is expected between the values for $R_e$ and $R_\mu$, and the value for $R_\tau$, even under the assumption of lepton universality [11].

Due to the small discrepancies between the values of $R_e$, $R_\mu$ and $R_\tau$, we use the lepton universality rule to define the $R_l$ [11]:

$$R_l := \Gamma_{had} / \Gamma_{lept} = 20.765 \pm 0.026 \tag{1.12}$$

We know:

$$\Gamma_{tot} = \Gamma_{had} + 3\Gamma_{lept} + N_\nu \Gamma_\nu \tag{1.13}$$

and we define:

$$\Gamma_{inv} = N_\nu \Gamma_\nu. \tag{1.14}$$

Where *inv* means invisible.

From (1.6), we have:

$$\Gamma_{tot}^2 = 12\pi \cdot \frac{\Gamma_{lept} \cdot \Gamma_{had}}{M_Z^2 \sigma_0}. \tag{1.15}$$

We notice again that here $\hbar = c = 1$. From the (1.15) we can do the following calculation:

$$\frac{\Gamma_{tot}^2}{\Gamma_{lept}^2} = 12\pi \cdot \frac{\Gamma_{lept} \cdot \Gamma_{had}}{\Gamma_{lept}^2 M_Z^2 \sigma_0}$$

$$\Rightarrow \quad \frac{\Gamma_{tot}^2}{\Gamma_{lept}^2} = 12\pi \cdot \frac{\Gamma_{had}}{\Gamma_{lept} M_Z^2 \sigma_0} \stackrel{(1.12)}{=} 12\pi \cdot \frac{R_\ell}{M_Z^2 \sigma_0}$$

$$\Rightarrow \quad \frac{\Gamma_{tot}}{\Gamma_{lept}} = \sqrt{\frac{12\pi R_\ell}{M_Z^2 \sigma_0}} \tag{1.16}$$

Now from (1.14) we can reach our goal, namely the neutrino species $N_\nu$:

$$N_\nu = \frac{\Gamma_{inv}}{\Gamma_\nu} \stackrel{(1.13)\&(1.14)}{=} \frac{(\Gamma_{tot} - \Gamma_{had} - 3\Gamma_{lept})}{\Gamma_\nu}$$

$$\Rightarrow \quad N_\nu = \frac{\Gamma_{lept}}{\Gamma_{lept}} \cdot \frac{(\Gamma_{tot} - \Gamma_{had} - 3\Gamma_{lept})}{\Gamma_\nu} = \frac{\Gamma_{lept}}{\Gamma_\nu} \cdot \frac{(\Gamma_{tot} - \Gamma_{had} - 3\Gamma_{lept})}{\Gamma_{lept}}$$

$$\Rightarrow \qquad N_\nu = \frac{\Gamma_{lept}}{\Gamma_\nu} \cdot \left( \frac{\Gamma_{tot}}{\Gamma_{lept}} - \frac{\Gamma_{had}}{\Gamma_{lept}} - \frac{3\Gamma_{lept}}{\Gamma_{lept}} \right) \stackrel{(1.12)}{=} \frac{\Gamma_{lept}}{\Gamma_\nu} \cdot \left( \frac{\Gamma_{tot}}{\Gamma_{lept}} - R_\ell - 3 \right)$$

$$\Rightarrow \qquad N_\nu \stackrel{(1.16)}{=} \frac{\Gamma_{lept}}{\Gamma_\nu} \cdot \left( \sqrt{\frac{12\pi R_\ell}{M_Z^2 \sigma_0}} - R_\ell - 3 \right). \tag{1.17}$$

In the Standard Model, the ratio $\Gamma_{lept}/\Gamma_\nu$ is very accurately known [1]:

$$\left. \frac{\Gamma_{lept}}{\Gamma_\nu} \right|_{SM} = 1.991 \pm 0.001. \tag{1.18}$$

Using this value along with the experimentally determined Z-boson mass $M_Z$ = (91.1867 ±0.0021) GeV, $\sigma_o$ = ( 41.491 ± 0.058 ) nb , $R_l$ = 20.765 ±0.026 and $\Gamma_{lept}$ = (83.90 ±0.10 ) MeV measured at LEP [11] in the equation (1.17), gives:

$$N_\nu = 2.994 \pm 0.011; \quad \Gamma_{inv} = (500.1 \pm 1.9)\text{ MeV}. \tag{1.19}$$

From (1.17) we can see $N_\nu \propto 1/\sqrt{\sigma_0}$. So it follows, then that when $N_\nu \nearrow$, $\sigma_o$ should be $\searrow$, similarly when $N_\nu \searrow$ then $\sigma_o$ should be $\nearrow$. Therefore, we can now demonstrate that there are 3 neutrino species when $m_\nu \leq m_Z/2$, with $m_Z$ = (91.1867 ±0.0021) GeV.

## 2  Dirac and Majorana masses

Before we go on to more issues, it's useful to examine in a bit of detail the possible patterns of neutrino masses. First, we want to review how fermion masses originate in the field theory. We consider a spinor field $\psi$ with the Dirac Lagrangian:

$$L_{Dirac} = i\overline{\psi}\gamma^\mu \partial_\mu \psi - m\overline{\psi}\psi. \tag{2.1}$$

So the mass term is:

$$L_{mass} = -m\overline{\psi}\psi = -m(\overline{\psi}_L \psi_R + \overline{\psi}_R \psi_L). \tag{2.2}$$

Just like normal, $\overline{\psi} = \psi^+ \gamma^0$, $\gamma^5 = i\gamma^0 \gamma^1 \gamma^2 \gamma^3$, and the projections:

$$\psi_L = \frac{1}{2}(1-\gamma^5)\psi; \quad \psi_R = \frac{1}{2}(1+\gamma^5)\psi. \tag{2.3}$$

The inspection of the Lagrangian (2.1) shows that it is invariant under the phase transformation [5]

$$\psi(x) \to e^{i\alpha}\psi(x). \tag{2.4}$$

Where $\alpha$ is a real constant. For $\overline{\psi}(x)$:

$$\overline{\psi}(x) \to e^{-i\alpha}\overline{\psi}(x). \tag{2.5}$$

The family of phase transformations $U(\alpha) := e^{i\alpha}$, where a single parameter $\alpha$ may run continuously over real numbers, forms a unitary Abelian group known as U(1) group.

Obviously it conserves the fermion number and gives equal mass for particles and antiparticles [1], it means:

$$m = m_{\overline{\psi}} = m_{\psi}. \tag{2.6}$$

Since the $\gamma^5$ is diagonal in the Weyl representation, so we choose to apply it here:

$$\gamma^0 = \begin{pmatrix} 0 & 1 \\ 1 & 0 \end{pmatrix}; \quad \gamma^i = \begin{pmatrix} 0 & \sigma^i \\ -\sigma^i & 0 \end{pmatrix}, i = 1,2,3 \tag{2.7}$$

and

$$\gamma^5 = i\gamma^0\gamma^1\gamma^2\gamma^3 = \begin{pmatrix} -1 & 0 \\ 0 & 1 \end{pmatrix}. \tag{2.8}$$

We notice that all these are 4×4 matrices. Of these the pauli Matrices are:

$$\sigma^1 = \begin{pmatrix} 0 & 1 \\ 1 & 0 \end{pmatrix}, \quad \sigma^2 = \begin{pmatrix} 0 & -i \\ i & 0 \end{pmatrix}, \quad \sigma^3 = \begin{pmatrix} 1 & 0 \\ 0 & -1 \end{pmatrix}. \tag{2.9}$$

The 4-component Dirac spinor $\psi$ is:

$$\psi = \begin{pmatrix} \xi_a \\ \dot{\chi}^a \end{pmatrix}. \tag{2.10}$$

We now perform the following calculations:

$$\overline{\psi} = \psi^+\gamma^0 = \begin{pmatrix} \xi_a \\ \dot{\chi}^a \end{pmatrix}^+ \gamma^0 = \begin{pmatrix} \xi_a^* & \dot{\chi}^{a*} \end{pmatrix} \begin{pmatrix} 0 & 1 \\ 1 & 0 \end{pmatrix} = \begin{pmatrix} \dot{\chi}^{a*} & \xi_a^* \end{pmatrix} \tag{2.11}$$

$$\psi_L = \frac{1}{2}(1-\gamma^5)\psi = \frac{1}{2}\left(1-\begin{pmatrix} -1 & 0 \\ 0 & 1 \end{pmatrix}\right)\begin{pmatrix} \xi_a \\ \dot{\chi}^a \end{pmatrix} = \frac{1}{2}\begin{pmatrix} 2 & 0 \\ 0 & 0 \end{pmatrix}\begin{pmatrix} \xi_a \\ \dot{\chi}^a \end{pmatrix}$$

$$\Rightarrow \quad \psi_L = \begin{pmatrix} 1 & 0 \\ 0 & 0 \end{pmatrix}\begin{pmatrix} \xi_a \\ \dot{\chi}^a \end{pmatrix} = \begin{pmatrix} \xi_a \\ 0 \end{pmatrix} \tag{2.12}$$

$$\overline{\psi}_L = \psi_L^+ \gamma^0 = \begin{pmatrix} \xi_a \\ 0 \end{pmatrix}^+ \gamma^0 = \begin{pmatrix} \xi_a^* & 0 \end{pmatrix} \begin{pmatrix} 0 & 1 \\ 1 & 0 \end{pmatrix} = \begin{pmatrix} 0 & \xi_a^* \end{pmatrix} \quad (2.13)$$

One can show it easily that:

$$\overline{\psi}_L = \psi_L^+ \gamma^0 = \overline{\psi} \frac{1}{2}(1+\gamma^5) = \begin{pmatrix} 0 & \xi_a^* \end{pmatrix} \quad (2.14)$$

the analog to our result is as follows:

$$\psi_R = \frac{1}{2}(1+\gamma^5)\psi = \begin{pmatrix} 0 \\ \dot{\chi}^a \end{pmatrix}; \quad (2.15)$$

$$\overline{\psi}_R = \psi_R^+ \gamma^0 = \overline{\psi} \frac{1}{2}(1-\gamma^5) = \begin{pmatrix} \dot{\chi}^{a*} & 0 \end{pmatrix}. \quad (2.16)$$

Using these equations, it is easy to see that the Dirac mass term connects $\xi$ with $\dot{\chi}$ [1]. So we have:

$$L_{Dirac} \stackrel{(2.2)}{=} -m_D(\overline{\psi}_L \psi_R + \overline{\psi}_R \psi_L) = -m_D(\xi_a^* \dot{\chi}^a + \dot{\chi}^{a*} \xi_a). \quad (2.17)$$

We know that dotted spinors are related to the complex conjugate of an undotted spinor $\dot{\xi} \sim \xi^*$ [1]. Choosing a phase convention where

$$\xi_a^* = \dot{\xi}_a; \quad \dot{\chi}^{a*} = \chi^a \quad (2.18)$$

one can rewrite the equation (2.17) as the following:

$$L_{Dirac} = -m_D(\dot{\xi}_a \dot{\chi}^a + \chi^a \xi_a). \quad (2.19)$$

Particles carry any U(1) quantum number, like electromagnetic charge, (for instance e⁻), must preserve this quantum number, and therefore the $L_{Dirac}$ is the only possible mass term. However neutrinos don't have electromagnetic charge, so they may have some other mass term that violates the fermion number (lepton number). The other choice for the neutrino is Majorana mass. One note that the antiparticle of Majorana particle is the Majorana particle itself. This is actually a very good feature for massive neutrinos. We note that handedness is not a good quantum number, as can be shown easily that $\gamma^5$ does not commute with the Hamiltonian. We see that the helicity is conserved, by "overtaking" the particle that concerned, but the helicity is frame dependent [5]. So how can we have a massive neutrino and still ensure that weak interactions couple only to $\nu_L$ and $\overline{\nu}_R$? It is the Majorana neutrino. We note here, that the neutrino is its own antiparticle. So we may identify $\nu_L$ and $\overline{\nu}_R$ as two helicity components of a four-component spinor. Then the other two neutrinos $\nu_R$ and $\overline{\nu}_L$ (if they exist) will have a different mass. One can define a four-component Majorana spinor with Weyl spinor $\xi$ and its complex conjugate $\dot{\xi}$:

$$\psi_M = \begin{pmatrix} \xi_a \\ \dot{\xi}^a \end{pmatrix} \qquad (2.20)$$

One can directly have the chiral projections of $\psi_M$ like this:

$$(\psi_M)_L = \frac{1}{2}(1-\gamma^5)\psi_M = \begin{pmatrix} \xi_a \\ 0 \end{pmatrix}; \qquad (2.21)$$

$$\overline{(\psi_M)_L} = \overline{\psi_M}\frac{1}{2}(1+\gamma^5) = \begin{pmatrix} 0 & \xi_a \end{pmatrix}; \qquad (2.22)$$

$$(\psi_M)_R = \frac{1}{2}(1+\gamma^5)\psi_M = \begin{pmatrix} 0 \\ \dot{\xi}^a \end{pmatrix}; \qquad (2.23)$$

$$\overline{(\psi_M)_R} = \overline{\psi_M}\frac{1}{2}(1-\gamma^5) = \begin{pmatrix} \dot{\xi}^{a*} & 0 \end{pmatrix} = \begin{pmatrix} \xi^a & 0 \end{pmatrix}. \qquad (2.24)$$

One see, it's not independent, and we can construct the $(\psi_M)_R$ by using the charge conjugate matrix $\widetilde{C}$. In the Weyl basis is $\widetilde{C}$ given as following [1]:

$$\widetilde{C} = \begin{bmatrix} \varepsilon_{ab} & \\ & \varepsilon^{ab}_{\dot{b}} \end{bmatrix} = \begin{bmatrix} 0 & -1 & 0 & 0 \\ 1 & 0 & 0 & 0 \\ 0 & 0 & 0 & 1 \\ 0 & 0 & -1 & 0 \end{bmatrix}. \qquad (2.25)$$

More generally is $\widetilde{C}$ given by:

$$\widetilde{C} = C\gamma^{0T} \qquad (2.26)$$

And the matrix $C$ is connected with how Dirac fields transform under charge conjugation. A Dirac field $\psi$ is transformed into its Hermitian conjugate $\psi^+$ under the charge conjugation operator $U(C)$ like following [1]:

$$U(C)\psi U(C)^{-1} = C\psi^+, \qquad (2.27)$$

and $C$ obeys the restriction [1]:

$$C\gamma_\mu^* C^{-1} = -\gamma_\mu \qquad (2.28)$$

In general it's depend on the $\gamma$-matrix basis that used. One can easily show, that:

$$(\psi_M)_R = \begin{pmatrix} 0 \\ \dot{\xi}^a \end{pmatrix} = \begin{pmatrix} 0 \\ \varepsilon^{ab}\dot{\xi}_b \end{pmatrix} = \widetilde{C}\overline{(\psi_M)_L}^T \ ; \tag{2.29}$$

$$\overline{(\psi_M)_R} = \begin{pmatrix} \xi^a & 0 \end{pmatrix} = \begin{pmatrix} \varepsilon^{ab}\xi_b & 0 \end{pmatrix} = (\psi_M)_L^T \widetilde{C} \tag{2.30}$$

The Majorana mass term is:

$$L_{Majorana} = -\frac{1}{2}m_M \overline{\psi_M}\psi_M = -\frac{1}{2}m_M(\overline{(\psi_M)_L}(\psi_M)_R + \overline{(\psi_M)_R}(\psi_M)_L) \ ; \tag{2.31}$$

$$\Rightarrow \qquad L_{Majorana} = -\frac{1}{2}m_M(\dot{\xi}_a\dot{\xi}^a + \xi^a\xi_a). \tag{2.32}$$

Using the equations (2.29), (2.30), we can write the (2.32) in an other manner, which has only $(\psi_M)_L$ or $(\psi_M)_R$ term:

$$L_{Majorana} = -\frac{1}{2}m_M\left(\overline{(\psi_M)_L}\widetilde{C}\overline{(\psi_M)_L}^T + (\psi_M)_L^T \widetilde{C}(\psi_M)_L\right); \tag{2.33}$$

$$L_{Majorana} = -\frac{1}{2}m_M\left((\psi_M)_R^T \widetilde{C}(\psi_M)_R + \overline{(\psi_M)_R}\widetilde{C}\overline{(\psi_M)_R}^T\right). \tag{2.34}$$

Now we summarize the all possible mass terms for neutrinos like follows:

$$L_{mass}^\nu = -\left[\overline{\nu_R}m_D\nu_L + \overline{\nu_L}m_D^+\nu_R\right] - \frac{1}{2}\left[\overline{\nu_R}\widetilde{C}m_S\overline{\nu_R}^T + \nu_R^T\widetilde{C}m_S^+\nu_R\right] \\ - \frac{1}{2}\left[\nu_L^T\widetilde{C}m_T\nu_L + \overline{\nu_L}\widetilde{C}m_T^+\overline{\nu_L}^T\right] \tag{2.35}$$

We note that for our convenience we write here the spinor $\nu$ instead of $\psi$. Here the $m_D$ conserves the lepton number and often called Dirac mass. Both $m_S$ and $m_T$ violate the lepton number, these are Majorana masses.

## 3 The See-Saw mechanism

At first we introduce the charge conjugate field $\psi^c$, it's defined as following:

$$\psi^c(x) = C\psi^{+T}(x) \tag{3.1}$$

Use $\overline{\psi} = \psi^+\gamma^0$ and $(\gamma^0)^2 = 1$, we have:

$$\psi^c(x) = C\gamma^{0T}\gamma^{0T}\psi^{+T}(x) = C\gamma^{0T}\overline{\psi}^T(x) = \widetilde{C}\overline{\psi}^T(x) \tag{3.2}$$

with

$$\psi^c = \widetilde{C}\overline{\psi}^T; \qquad \overline{\psi^c} = \psi^T \widetilde{C} \qquad (3.3)$$

and $\widetilde{C}\widetilde{C} = -1$ we can write:

$$\overline{\nu_R}\nu_L = -\nu_L^T \overline{\nu_R}^T = \nu_L^T \widetilde{C}\widetilde{C}\overline{\nu_R}^T = \overline{\nu_L^c}\nu_R^c. \qquad (3.4)$$

So we can write $\overline{\nu_R}\nu_L$ as following:

$$\overline{\nu_R}\nu_L = \frac{1}{2}\left[\overline{\nu_R}\nu_L + \overline{\nu_L^c}\nu_R^c\right] \qquad (3.5)$$

Then we can rewrite the equation (2.35) with another way, just like following:

$$L_{mass}^\nu = -\frac{1}{2}\left[\begin{pmatrix}\overline{\nu_L^c} & \overline{\nu_R}\end{pmatrix}\begin{pmatrix}m_T & m_D^T \\ m_D & m_S\end{pmatrix}\begin{pmatrix}\nu_L \\ \nu_R^c\end{pmatrix}\right] + h.c. \qquad (3.6)$$

We note that the notation *h.c.* denotes the Hermitian conjugate. Now we define M as following:

$$M = \begin{pmatrix}m_T & m_D^T \\ m_D & m_S\end{pmatrix} \qquad (3.7)$$

Since we have found that there are 3 species of neutrinos, therefore the $m_D$, $m_T$ and $m_S$ are each $3\times 3$ matrix, and $M$ is a $6\times 6$ Matrix. For the 3 generation of neutrinos, the 6 eigenstates $m_i$ are the eigenvalue of $M$.

Since the $M$ is in general not diagonal, so we often diagonalize the $M$ with two unitary matrices $U_L$ and $U_R$:

$$M_{diag} = U_R^+ M U_L \qquad (3.8)$$

Actually in most case is $U_L = U_R$ and $U$ is real number, therefore in most of case $U^+$ will be then $U^T$.

Hence now we have a new Matrix $M_{diag}$. we introduce a new set of mass eigenfields $N_L$ and $N_R$ as:

$$\psi_L = U_L N_L; \qquad \psi_R = U_R N_R. \qquad (3.10)$$

Which the $\psi_L$ and $\psi_R$ are defined as following:

$$\psi_L = \begin{pmatrix}\nu_L \\ \nu_R^c\end{pmatrix}; \qquad \psi_R = \begin{pmatrix}\nu_L^c \\ \nu_R\end{pmatrix} \qquad (3.11)$$

We consider here a physically very interesting and relatively simple case. Now we think about an one dimensional case, it means that the $m_D$, $m_T$ and $m_S$ are each a scalar, thus is just only one family of neutrinos. And we suppose that $m_T = 0$ and $m_S \gg m_D$ [1]. We notice that all elements here are real number and one-dimensional. Then we can write the $6\times 6$ Matrix $M$ simply as the following:

$$M = \begin{pmatrix} 0 & m_D \\ m_D & m_S \end{pmatrix}. \tag{3.12}$$

After a small calculation, we have two eigenvalues for the matrix $M$, it is:

$$\left\{ -\left( \frac{\sqrt{4 \cdot m_D^2 + m_S^2} - m_S}{2} \right), \quad \frac{\sqrt{4 \cdot m_D^2 + m_S^2} + m_S}{2} \right\} \tag{3.13}$$

Since we assume that $m_S \gg m_D$, the right term of (3.13) is approximately equal to $(m_S + m_S)/2 = m_S$, for the left term we can rewrite it as:

$$-\left( \frac{\sqrt{4 \cdot m_D^2 + m_S^2} - m_S}{2} \right) = -\left( \frac{m_S \sqrt{4 \cdot \frac{m_D^2}{m_S^2} + 1} - m_S}{2} \right)$$

$$\underset{\left(\frac{m_D^2}{m_S^2} \ll 1\right)}{\approx} -\left( \frac{m_S \left(1 + \frac{\left(4 \cdot \frac{m_D^2}{m_S^2}\right)}{2}\right) - m_S}{2} \right) = -m_S \cdot \frac{m_D^2}{m_S^2}$$

$$= -\frac{m_D^2}{m_S} \tag{3.14}$$

In this case, we see that the spectrum of the neutrino mass splits into a very light neutrino mass $m_D^2/m_S$ and a very heavy one $m_S$. One note these are approximately. This mechanism is so called seesaw mechanism. It's natural to expect that $m_D$ should be of the order of the charged lepton mass, corresponding to the neutrino in our case: $m_D \sim m_l$ [1]. Then we can easily see:

$$(m_\nu)_{light} \sim m_\ell \left(\frac{m_\ell}{m_S}\right) << m_\ell << (m_\nu)_{heavy} \sim m_S \qquad (3.15)$$

It is very helpful for us to understand why neutrinos could be much lighter than the corresponding charged leptons, when there exist a very heavy mass that associates with the right-handed neutrinos. We note also that since the right-handed neutrinos are a $SU(2) \times U(1)$ singlet, there is no constrain for the mass scale $m_S$. Actually this is a very important and useful feature of the seesaw mechanism. As we will see in Chapter 5, because of the cosmological constrains, the neutrino mass should be less than about 10 eV and in principle, cosmology allows also the neutrinos to exist with mass greater than, say about 10 GeV. So we expect that in a sooner future, we can have more evidence about the existence of the heavy neutrinos, self-evident these are massive neutrinos.

We have shown the eigenvalue of the *M*, certainly we want now to diagonalize the Matrix *M*, just like before we do it with the approximately way, and we have a orthogonal matrix *U*:

$$U = \begin{pmatrix} 1 & \frac{m_D}{m_S} \\ -\frac{m_D}{m_S} & 1 \end{pmatrix} \qquad (3.16)^*$$

So from the definition (3.10) and we use the fact $U_L = U_R$ along with the orthogonality of the *U* we have:

$$N_L = U_L^T \psi_L = \begin{pmatrix} 1 & -\frac{m_D}{m_S} \\ \frac{m_D}{m_S} & 1 \end{pmatrix} \begin{pmatrix} \nu_L \\ \nu_R^c \end{pmatrix} := \begin{pmatrix} N_1 \\ N_2 \end{pmatrix}_L ; \qquad (3.17)$$

$$N_R = U_R^T \psi_R = \begin{pmatrix} 1 & -\frac{m_D}{m_S} \\ \frac{m_D}{m_S} & 1 \end{pmatrix} \begin{pmatrix} \nu_L^c \\ \nu_R \end{pmatrix} := \begin{pmatrix} N_1 \\ N_2 \end{pmatrix}_R . \qquad (3.18)$$

We note that the eigenstates $N_1$ and $N_2$ are self-conjugate, so they are Majorana states. We see the following:

---

* One does a simple calculation and has the following result: $U^T M U = \begin{pmatrix} -\frac{m_D^2}{m_S} & -\frac{m_D^3}{m_S^2} \\ -\frac{m_D^3}{m_S^2} & 2\frac{m_D^2}{m_S} + m_S \end{pmatrix}$.

$$N_1 = N_{1L} + N_{1R} = (\nu_L + \nu_L^c) - \frac{m_D}{m_S}(\nu_R^c + \nu_R) = N_1^c \qquad (3.19)$$

$$N_2 = N_{2L} + N_{2R} = (\nu_R^c + \nu_R) + \frac{m_D}{m_S}(\nu_L + \nu_L^c) = N_2^c \qquad (3.20)$$

Because only the link-handed neutrinos enter in the weak interactions, and $m_D << m_S$, so we see, for the practical purpose the essentially term is only $N_{1L}$, and obviously it associated with the light neutrino eigenstate. We follow again the definition (3.10) and use the fact $U_L = U_R$:

$$\psi_L = \begin{pmatrix} \nu_L \\ \nu_R^c \end{pmatrix} = UN_L = \begin{pmatrix} 1 & \frac{m_D}{m_S} \\ -\frac{m_D}{m_S} & 1 \end{pmatrix} \begin{pmatrix} N_{1L} \\ N_{2L} \end{pmatrix} = \begin{pmatrix} N_{1L} + \frac{m_D}{m_S} N_{2L} \\ -\frac{m_D}{m_S} N_{1L} + N_{2L} \end{pmatrix} \qquad (3.21)$$

So we have:

$$\nu_L = N_{1L} + \frac{m_D}{m_S} N_{2L} \qquad (3.22)$$

Analog to the link-handed neutrinos, the essentially term for right-handed neutrinos is $N_{2R}$, the heavy eigenstate. We have now:

$$\nu_R = N_{2R} - \frac{m_D}{m_S} N_{1R} \qquad (3.23)$$

We note again, because the Matrix $U$ is only approximately diagonalize the $M$, so all result here are approximately.

This one dimensional Result can be easily expand to the 3 dimensional case with 3 generations of neutrinos, now we have:

$$M = \begin{pmatrix} 0 & m_D^T \\ m_D & m_S \end{pmatrix} \qquad (3.24)$$

Here the $m_D$ and $m_S$ are each $3 \times 3$ matrix, and $M$ is a $6 \times 6$ Matrix. Again the spectrum separates into a light and heavy neutrino sector, which as follows:

$$(M_\nu)_{light} = m_D^T m_s^{-1} m_D \qquad (3.25)$$

$$(M_\nu)_{heavy} = m_S . \qquad (3.26)$$

## 4    Neutrino oscillations

In the last few chapters we have discussed some theoretical possibility of the neutrino mass, but how can we find whether neutrino has or not mass? A classical way by measure beta decay is discussed in the next section. However all experiments have measured almost a negative mass $m_{\nu_e}$. The most popular way now to determine the neutrino mass is the so-called neutrino oscillation, that is: we suppose neutrinos are massive and mixed. This is very similar to that of the oscillation in neutral kaons (CP violation). This will be discussed in section 4.2 and the following. However also in this case we still don't have the definitively answer for the neutrino mass.

## 4.1 Direct Mass Measurements

In this method we look at a nuclear beta-decay process:

$$N_i(A,Z) \rightarrow N_f(A,Z+1) + e^- + \bar{\nu}_e \qquad (4.1)$$

A... Mass number
N... Charge number
i... Initial
f... Final

So what we want is to study the influence of neutrino mass on the beta-decay process. Now we define a quantity $K(E)$, after some calculations, one has the following relation near the end point $E = E_0$ [9]:

$$K(E) := \left(\frac{d\Gamma/dE}{p \cdot E \cdot F(Z,E)}\right)^{1/2} \sim \left((E_0 - E)\sqrt{(E_0 - E)^2 - m_\nu^2}\right)^{1/2} \qquad (4.2)$$

F(E,Z)... Coulomb function (Fermi function)
Γ... Total decay rate
p... magnitude of the electron momentum (scalar)
E... Total electron energy
$E_0$... Total energy released

Actually this is the most useful quantity for measuring the neutrino mass. The plot of this function is so-called Kurie plot. For $m_\nu = 0$, the Kurie plot becomes a straight line $\sim E_0 - E$, intercepting the energy axis at $E = E_0$. This is often the most accurate way to measure the $E_0$ or Q value [9], which defined as follows:

$$Q = E_0 - m_e \qquad (4.3)$$

This is the maximal kinetic energy of the electron. Otherwise when $m_\nu \neq 0$, the spectrum shape near the end point $E = E_0$ will be no more a straight line. We see the deviation is most sensitive to $m_\nu$ when the value of $E_0$ or say Q is small. The Kurie plot is shown in Fig. (4.1), one note that for the illustrative purpose, it's exaggerated. Among all the known beta-decays, the Tritium beta-decay is most widely used to search for the mass of $\bar{\nu}_e$. This is because the process:

$$^3H \rightarrow {}^3He + e^- + \bar{\nu}_e \qquad (4.4)$$

has the smallest Q value, namely Q = 18.6 KeV. Tritium experiments are sensitive to neutrino mass in the "few eV" range [1]. The table (4.1) shows the recently neutrino mass limits from Tritium beta-decay [1]. The reason for the minus sign here is unfortunately still unknown. However, very recently, the Troitsk experiment has been able to determine a very stringent result for the largest eigenvalue "$m_{\nu_e}$" with 90% C.L. (Confidence Level) [1]. Similar, but less accurate, we also have got a upper limit for $m_{\nu_\mu}$ and $m_{\nu_\tau}$ at PSI and LEP [1]:

$$"m_{\nu_\mu}^2" = (-0.016 \pm 0.028) MeV^2 ; \quad "m_{\nu_\mu}" < 170 \text{ KeV } (90\% \text{ C.L.}) \quad (4.5)$$

$$"m_{\nu_\tau}" < 18.2 \text{ MeV } (95\% \text{ C.L.}) \quad (4.6)$$

**TABLE (4.1)** neutrino mass limit from Tritium beta-decay, from Ref. [1].

| Experiment | "$m_{\nu_e}^2$" ($eV^2$) | | | "$m_{\nu_\tau}$" ($eV$) |
|---|---|---|---|---|
| Tokyo      | -65  | ± 85  | ± 65 | < 13.1 |
| Los Alamos | -147 | ± 68  | ± 41 | < 9.3  |
| Zürich     | -24  | ± 48  | ± 61 | < 11.7 |
| Livermore  | -130 | ± 20  | ± 15 | < 7.0  |
| Mainz      | -22  | ± 17  | ± 14 | < 5.6  |
| Troitsk    | 1.5  | ± 5.9 | ± 3.6 | < 3.9 |

As we will see in Chapter 5, because of the cosmological constrains, the neutrino mass should be less than about 10 eV, so in my view, the (4.5) and (4.6) are more or less caused from the measurement errors.

**4.2  Neutrino oscillations in vacuum**

If neutrinos are massive, there is no reason to believe the weak interaction eigenstates and the mass eigenstates are identical [3]. The consequence of neutrino mixing, first suggested by Pontecorvo is called neutrino oscillation [8]. When it's true, that means: The so-called $\nu_l$, (*l* denotes Lepton) neutrinos (weak interaction eigenstates) that created in the experiments are generally not a physical particle, but rather a superposition of the physical fields $\nu_\alpha$ with different masses $m_\alpha$ (mass eigenstates) :

$$|\nu_\ell\rangle = \sum_\alpha U_{\ell\alpha} |\nu_\alpha\rangle \quad (4.7)$$

Where *U* is a unitary matrix. Since the most of the data presented usually 2 flavors of neutrinos, we will discuss the oscillations also in this simple case. In case of need we can generalize the formalism straightforwardly to 3 generations neutrinos. For the 2 flavors of neutrinos we have the following relation between the weak interaction eigenstates and the mass eigenstates:

$$\begin{cases} |\nu_e\rangle = \cos\theta_V |\nu_1\rangle + \sin\theta_V |\nu_2\rangle \\ |\nu_\mu\rangle = -\sin\theta_V |\nu_1\rangle + \cos\theta_V |\nu_2\rangle \end{cases} \quad (4.8)$$

Where $\theta_V$ denotes the vacuum mixing angle. For our future convenience, we develop this equation a little bit more. We can rewrite it as a matrix form:

$$\begin{bmatrix} v_e \\ v_\mu \end{bmatrix} = U \begin{bmatrix} v_1 \\ v_2 \end{bmatrix}, \quad where \quad U = \begin{pmatrix} \cos\theta_V & \sin\theta_V \\ -\sin\theta_V & \cos\theta_V \end{pmatrix}. \tag{4.9}$$

One can easily check the orthogonality of the Matrix $U$. More generally we say it's unitary. Then we have:

$$\begin{bmatrix} v_1 \\ v_2 \end{bmatrix} = U^{-1} \begin{bmatrix} v_e \\ v_\mu \end{bmatrix} = U^+ \begin{bmatrix} v_e \\ v_\mu \end{bmatrix} = \begin{pmatrix} \cos\theta_V & -\sin\theta_V \\ \sin\theta_V & \cos\theta_V \end{pmatrix} \begin{bmatrix} v_e \\ v_\mu \end{bmatrix} \tag{4.10}$$

$$\Rightarrow \quad \begin{cases} |v_1\rangle = \cos\theta_V |v_e\rangle - \sin\theta_V |v_\mu\rangle \\ |v_2\rangle = \sin\theta_V |v_e\rangle + \cos\theta_V |v_\mu\rangle \end{cases}. \tag{4.11}$$

For the time evolution of the states we need to solved Schrödinger equation (In this case this is sometimes known as the Wigner-Weisskopf equation.):

$$i\frac{\partial}{\partial t}\begin{bmatrix} v^1 \\ v^2 \end{bmatrix} = H \begin{bmatrix} v^1 \\ v^2 \end{bmatrix}, \tag{4.12}$$

with H for vacuum oscillations with the eigenstates $v_1$ and $v_2$ is here given by

$$H = \begin{pmatrix} E_1 & 0 \\ 0 & E_2 \end{pmatrix}. \tag{4.13}$$

Where

$$E_i = \sqrt{\vec{p}^2 + m_i^2}. \tag{4.14}$$

One note that the 3-components momentum $\vec{p}$ of the different components in the neutrinos beam are the same and are determined by momentum conservation in the process in which the neutrinos are created [8]. Another important thing is that: The $H$ is only diagonal for the eigenstates $v_1$ and $v_2$. We will discuss it more lately. After solved the equation (4.9), we have the solution:

$$|v^i(t)\rangle = e^{-iE_i t} |v^i(0)\rangle, \quad i = 1, 2; \tag{4.15}$$

When $m_1 \neq m_2$, one can easily see that: when we have the weak interaction eigenstates $v_e$ in $t = 0$, then it will be evolved into a superposition of $v_e$ and $v_\mu$ states. We suppose at $t = 0$, we have the state $v_e$, and use the equations (4.15), (4.8) and (4.11), now we have the time evolution of the state $v_e(t)$ as following:

$$|v_e(t)\rangle = \cos\theta_V \cdot e^{-iE_1 t} |v_1\rangle + \sin\theta_V \cdot e^{-iE_2 t} |v_2\rangle$$

$$= \cos\theta_V \cdot e^{-iE_1 t}\left(\cos\theta_V |v_e\rangle - \sin\theta_V |v_\mu\rangle\right) + \sin\theta_V \cdot e^{-iE_2 t}\left(\sin\theta_V |v_e\rangle + \cos\theta_V |v_\mu\rangle\right)$$

$$= \left(\cos^2\theta_V \cdot e^{-iE_1 t} + \sin^2\theta_V \cdot e^{-iE_2 t}\right)|v_e\rangle + \left(\cos\theta_V \cdot \sin\theta_V \cdot \left(e^{-iE_2 t} - e^{-iE_1 t}\right)\right)|v_\mu\rangle$$

$$:= A_{ee}|v_e\rangle + A_{e\mu}|v_\mu\rangle. \quad (4.16)$$

So one can easily calculate the oscillation probabilities to the weak interaction states from the original state $v_e$ at time $t$:

$$P(v_e \to v_e; t) = |A_{ee}(t)|^2 = 1 - \frac{1}{2}\sin^2 2\theta_V [1 - \cos((E_2 - E_1)t)] \quad (4.17)$$

$$P(v_e \to v_\mu; t) = |A_{e\mu}(t)|^2 = \frac{1}{2}\sin^2 2\theta_V [1 - \cos((E_2 - E_1)t)] \quad (4.18)$$

We know that the neutrinos have a velocity near by the light velocity. Especially in our nature unit, is the light velocity $c = 1$, so the distance $L$ traveled by the neutrinos in a time t is about:

$$L \cong c \cdot t = t \quad (4.19)$$

Since the masses of neutrinos are small compared to the momentum, we have the following approximation for the equation (4.14):

$$E_i = \sqrt{\vec{p}^2 + m_i^2} = |\vec{p}|\sqrt{1 + \frac{m_i^2}{\vec{p}^2}}$$

$$\approx |\vec{p}|\left(1 + \frac{1}{2}\frac{m_i^2}{|\vec{p}|^2}\right) = |\vec{p}| + \frac{m_i^2}{2|\vec{p}|} \quad (4.20)$$

We define that

$$\Delta m^2 = m_2^2 - m_1^2, \quad (4.21)$$

and use the equation (4.20), and then we have:

$$E_2 - E_1 = \left(|\vec{p}| + \frac{m_2^2}{2|\vec{p}|}\right) - \left(|\vec{p}| + \frac{m_1^2}{2|\vec{p}|}\right) = \frac{\Delta m^2}{2|\vec{p}|} \quad (4.22)$$

We use the triangle form $\sin^2\frac{\alpha}{2} = \frac{1 - \cos\alpha}{2}$ along with the equations (4.18), (4.19), (4.20), (4.21) and (4.22), and then get:

Neutrino in Astrophysics and Cosmology                                                                - 17 -$$P(\nu_e \to \nu_\mu; t) = \frac{1}{2}\sin^2 2\theta_V \left[1 - \cos\left(\frac{\Delta m^2}{2|\vec{p}|}L\right)\right]$$

$$= \sin^2 2\theta_V \cdot \sin^2\left(\frac{\Delta m^2}{4|\vec{p}|}L\right). \tag{4.23}$$

Numerically it turns out that:

$$\frac{\Delta m^2}{4|\vec{p}|}L \approx 1.27 \frac{\Delta m^2(eV^2)}{|\vec{p}|(MeV)} L(m). \tag{4.24}$$

Then we have the probability $P(\nu_e \to \nu_\mu; t)$ as follows:

$$P(\nu_e \to \nu_\mu; t) = \sin^2 2\theta_V \cdot \sin^2\left(1.27 \frac{\Delta m^2(eV^2)}{|\vec{p}|(MeV)} L(m)\right). \tag{4.25}$$

We define an oscillation length as the following:

$$\frac{\Delta m^2}{2|\vec{p}|} L^{osc} = 2\pi, \tag{4.26}$$

and we get [9]:

$$L^{osc} = \frac{4\pi|\vec{p}|}{\Delta m^2} \approx \frac{2.48|\vec{p}|(MeV)}{\Delta m^2(eV^2)} Meter. \tag{4.27}$$

Obviously

$$\frac{\pi L}{L^{osc}} = \pi L \frac{1}{L^{osc}} = \pi L \frac{\Delta m^2}{4\pi|\vec{p}|} = \frac{\Delta m^2}{4|\vec{p}|} L. \tag{4.28}$$

Again we can rewrite the equation (4.23) as the following:

$$P(\nu_e \to \nu_\mu; t) = \sin^2 2\theta_V \cdot \sin^2\left(\frac{\pi L}{L^{osc}}\right). \tag{4.29}$$

One sees easily that:

$$P(\nu_e \to \nu_e; L) = 1 - P(\nu_e \to \nu_\mu; L) \tag{4.30}$$

We see that there are two factors which influence the oscillation probability. One is the mixing angle factor $\sin^2 2\theta_V$ and the second is the kinematical factor, it depends on the travel length $L$. Wee see from the equation (4.29), that the $L$ should be in the order $L^{osc}$ or greater.

When we see the equation (4.25), the factor $\sin^2\left(1.27\Delta m^2(eV^2)\frac{L(m)}{|\vec{p}|(MeV)}\right)$, we can see the sensitive for the mass difference is:

$$\Delta m^2 \approx \frac{|\vec{p}|(MeV)}{L(m)} eV^2 \approx \frac{E(MeV)}{L(m)} eV^2 \quad (4.31)$$

Here we use again the assume that $|\vec{p}| \gg m$ and we can write $E = \sqrt{\vec{p}^2 + m^2} \approx |\vec{p}|$.

We develop now the Schrödinger equation (4.12) a little bit more for the future use. Here we profit from the equations (4.9) and (4.10), which we have already developed. We use also the fact, that $U$ is time independent and unitary $(U^+ \cdot U = U \cdot U^+ = 1)$.

$$i\frac{\partial}{\partial t}\begin{bmatrix} v^1 \\ v^2 \end{bmatrix} = H_{diag}\begin{bmatrix} v^1 \\ v^2 \end{bmatrix}$$

$$\Rightarrow \quad i\frac{\partial}{\partial t}\left(U^+\begin{bmatrix} v^e \\ v^\mu \end{bmatrix}\right) = H_{diag}U^+\begin{bmatrix} v^e \\ v^\mu \end{bmatrix}$$

$$\Rightarrow \quad U^+ i\frac{\partial}{\partial t}\left(\begin{bmatrix} v^e \\ v^\mu \end{bmatrix}\right) = H_{diag}U^+\begin{bmatrix} v^e \\ v^\mu \end{bmatrix}$$

$$\Rightarrow \quad i\frac{\partial}{\partial t}\left(\begin{bmatrix} v^e \\ v^\mu \end{bmatrix}\right) = UH_{diag}U^+\begin{bmatrix} v^e \\ v^\mu \end{bmatrix} \quad (4.32)$$

We define $H := UH_{diag}U^+$.

$$\Rightarrow \quad i\frac{\partial}{\partial t}\left(\begin{bmatrix} v^e \\ v^\mu \end{bmatrix}\right) = H\begin{bmatrix} v^e \\ v^\mu \end{bmatrix} \quad (4.33)$$

So we have the solution:

$$\begin{bmatrix} v^e(t) \\ v^\mu(t) \end{bmatrix} = e^{-iHt}\begin{bmatrix} v^e(0) \\ v^\mu(0) \end{bmatrix} \quad (4.34)$$

We recall the MMP II proposition: If $U \in GL(n,C)$ and $A \in M_n(C)$, then is:

$$e^{UAU^{-1}} = Ue^A U^{-1} \tag{4.35}$$

So we can write:

$$e^{-iHt} = Ue^{-iH_{diag}t}U^{-1} \tag{4.36}$$

and we have:

$$A_{ee}(t) = \left[e^{-iHt}\right]_{11}; \quad A_{e\mu}(t) = \left[e^{-iHt}\right]_{12} \tag{4.37}$$

Analog to equation (4.16), we have:

$$|\nu_\mu(t)\rangle = A_{\mu e}(t)|\nu_e\rangle + A_{\mu\mu}(t)|\nu_\mu\rangle. \tag{4.38}$$

Then we have:

$$A_{\mu e}(t) = \left[e^{-iHt}\right]_{21}; \quad A_{\mu\mu}(t) = \left[e^{-iHt}\right]_{22} \tag{4.39}$$

Now we want to know, how does the $H$ explicit like? One recall the equation (4.20) $E_i \approx |\vec{p}| + \frac{m_i^2}{2|\vec{p}|}$ and (4.21) $\Delta m^2 = m_2^2 - m_1^2$. So we rewrite the $H_{diag} = \begin{pmatrix} E_1 & 0 \\ 0 & E_2 \end{pmatrix}$ as following:

$$H_{diag} \approx \left(|\vec{p}| + \frac{m_1^2 + m_2^2}{4|\vec{p}|}\right)\begin{pmatrix} 1 & 0 \\ 0 & 1 \end{pmatrix} + \frac{\Delta m^2}{4|\vec{p}|}\begin{pmatrix} -1 & 0 \\ 0 & 1 \end{pmatrix} \tag{4.40}$$

We notice that the first term in he equation (4.40) is irrelevant for us, because it is proportional to the unit matrix, and gives an overall phase factor, therefore we don't need it for the neutrino oscillation probabilities. Hence, effectively, instead of the $H_{diag}$, one can use $H_0$:

$$H_{diag} \to H_0 = -\frac{\Delta m^2}{4|\vec{p}|}\sigma_3 \tag{4.41}$$

One recalls the definition of the Pauli matrices in equation (2.9),

$$\sigma^1 = \begin{pmatrix} 0 & 1 \\ 1 & 0 \end{pmatrix}, \quad \sigma^2 = \begin{pmatrix} 0 & -i \\ i & 0 \end{pmatrix}, \quad \sigma^3 = \begin{pmatrix} 1 & 0 \\ 0 & -1 \end{pmatrix}. \tag{4.42}$$

So the Hamiltonian matrix $H_{vac}$ (which *vac* denotes vacuum) is like the following:

$$H_{vac} = U(\theta_V) H_0 U(\theta_V)^+ = \frac{\Delta m^2}{4|\vec{p}|} \begin{pmatrix} -\cos 2\theta_V & \sin 2\theta_V \\ \sin 2\theta_V & \cos 2\theta_V \end{pmatrix}$$

$$= \frac{\Delta m^2}{4|\vec{p}|} (\sin 2\theta_V \cdot \sigma_1 - \cos 2\theta_V \cdot \sigma_3). \tag{4.43}$$

### 4.3 Neutrino oscillations in matter

Wolfenstein pointed out that the patterns of neutrino oscillation might be significantly affected if the neutrinos propagate through a material medium rather than through the vacuum [8]. This effect is known as the MSW (Mikheyev-Smirnov-Wolfenstein) effect for the initials of the physicists who first discussed it [1] & [14]. The reason of the MSW effect is connected to the fact that there are no muons or taons in normal matter. It means the muon and taon neutrinos interact in matter only through the neutral current interactions, while the electron neutrinos interact in matter due to the both charged and neutral current interactions. Because of the charged current interactions, the electron neutrinos experience a slightly different index of refraction than the muon and taon neutrinos. With this different index of refraction for the electron neutrinos, it will change the time evolution of the system as it was in vacuum.

For a quantitative treatment of the above descriptions, we stick again to the simple case of two neutrino generations. In the last section we have developed the Hamiltonian matrix $H_{vac}$ in vacuum. Now we will do more developments for the neutrino oscillations in the matter. Generally we write that $H = H_{vac} + V$. Since we are only interested in the relative difference between the two neutrinos, we consider only the charged current interactions from the electron neutrinos. Therefore we must have the following form for the Hamiltonian:

$$H = H_{vac} + V_e \begin{pmatrix} 1 & 0 \\ 0 & 0 \end{pmatrix} \tag{4.44}$$

Here $V_e$ is a scalar, and the '$e$' denotes the charged current interactions for the electron neutrinos. One finds [14]

$$V_e = \pm\sqrt{2} G_F N_e \tag{4.45}$$

The sign depends on whether we deal with neutrinos (+) or antineutrinos (-). For our convenience we assume to have former. $G_F$ is the Fermi weak coupling constant and $N_e$ the electron density. So we have now:

$$H = H_{vac} + \sqrt{2} G_F N_e \begin{pmatrix} 1 & 0 \\ 0 & 0 \end{pmatrix}. \tag{4.46}$$

We emphasize again, that the overall phases are irrelevant for us. Therefore we subtract the equation (4.46) with a half identity as follows and use the equation (4.43):

$$H_{matter} = H_{vac} + \sqrt{2} G_F N_e \left( \begin{pmatrix} 1 & 0 \\ 0 & 0 \end{pmatrix} - \frac{1}{2} \begin{pmatrix} 1 & 0 \\ 0 & 1 \end{pmatrix} \right)$$

$$= H_{vac} + \frac{\sqrt{2} G_F N_e}{2} \begin{pmatrix} 1 & 0 \\ 0 & -1 \end{pmatrix}$$

$$= \frac{\Delta m^2}{4|\vec{p}|} \left( \sin 2\theta_V \cdot \sigma_1 - \cos 2\theta_V \cdot \sigma_3 \right) + \frac{G_F N_e}{\sqrt{2}} \cdot \sigma_3$$

$$= \frac{\Delta m^2}{4|\vec{p}|} \sin 2\theta_V \cdot \sigma_1 - \left( \frac{\Delta m^2}{4|\vec{p}|} \cos 2\theta_V - \frac{G_F N_e}{\sqrt{2}} \right) \cdot \sigma_3. \qquad (4.47)$$

Because of the new term in the equation (4.47), we have also new eigenstates for the $H_{matter}$. We call these $v_1^M$ and $v_2^M$. One has:

$$\begin{bmatrix} |v_e\rangle \\ |v_\mu\rangle \end{bmatrix} = \begin{pmatrix} \cos\theta_M & \sin\theta_M \\ -\sin\theta_M & \cos\theta_M \end{pmatrix} \begin{bmatrix} |v_1^M\rangle \\ |v_2^M\rangle \end{bmatrix} = U_M \begin{bmatrix} |v_1^M\rangle \\ |v_2^M\rangle \end{bmatrix} \qquad (4.48)$$

and

$$H_{matter} = U_M H_{matter}^{diag} U_M^+. \qquad (4.49)$$

So one gets the solution [1]:

$$H_{matter}^{diag} = -\sigma_3 \cdot \sqrt{\left( \frac{\Delta m^2}{4|\vec{p}|} \sin 2\theta_V \right)^2 - \left( \frac{\Delta m^2}{4|\vec{p}|} \cos 2\theta_V - \frac{G_F N_e}{\sqrt{2}} \right)^2} \qquad (4.50)$$

and the $\sin 2\theta_M$:

$$\sin 2\theta_M = \frac{\frac{\Delta m^2}{4|\vec{p}|} \sin 2\theta_V}{\sqrt{\left( \frac{\Delta m^2}{4|\vec{p}|} \sin 2\theta_V \right)^2 + \left( \frac{\Delta m^2}{4|\vec{p}|} \cos 2\theta_V - \frac{G_F N_e}{\sqrt{2}} \right)^2}} \qquad (4.51)$$

The term associated with the electron density gives a resonance phenomenon. This is very interesting for us; therefore we define a critical density $N_e^{crit}$ as the following:

$$N_e^{crit} = \frac{\Delta m^2 \cos 2\theta_V}{2\sqrt{2}|\vec{p}|G_F}. \tag{4.52}$$

We see at this electron density the $\sin 2\theta_M$ goes to 1; we say the matter mixing angle $\theta_M$ becomes maximal, irrespective of the vacuum mixing angle $\theta_V$. One can straightforwardly see that at this electron density the equation (4.50) reduces to:

$$H_{matter}^{diag}\bigg|_{N_e=N_e^{crit}} = -\left(\frac{\Delta m^2}{4|\vec{p}|}\sin 2\theta_V\right)\sigma_3 \tag{4.53}$$

In principle we can use the equation (4.23) for the oscillation probability. But we need do some changes on the equation. First of all, instead of $\theta_V$ in vacuum is the mixing angle now $\theta_M$. Second, since we have a different Hamiltonian, we should also replace the $\Delta m^2$ term with $\Delta m^2 \sin 2\theta$. We consider furthermore, that at the critical electron density, is $\sin 2\theta_M = 1$, so we get the following oscillation probability for the neutrinos in matter with the critical electron density:

$$P_{matter}(\nu_e \to \nu_\mu; t)\bigg|_{N_e=N_e^{crit}} = \sin^2 2\theta_M \cdot \sin^2\left(\frac{\Delta m^2 \sin 2\theta}{4|\vec{p}|}L\right)$$

$$= \sin^2\left(\frac{\Delta m^2}{4|\vec{p}|}\sin 2\theta \cdot L\right) \tag{4.54}$$

One sees from the equation (4.54), that we can get a full conversion from the electron neutrino weak interaction eigenstate into a muon neutrino weak interaction eigenstate, when we satisfy the following relation:

$$\frac{\Delta m^2}{4|\vec{p}|}\sin 2\theta \cdot L = \frac{(2n+1)\pi}{2}; \quad n = 0, 1, 2, \ldots \tag{4.55}$$

Another very interesting case is: The electron density is so large that the other terms in the Hamiltonian $H_{matter}^{diag}$ are negligible. In this situation one can have the Hamiltonian approximately as the following:

$$H_{matter}^{diag} = -\sigma_3 \cdot \frac{G_F N_e}{\sqrt{2}} \tag{4.56}$$

From the equation (4.51), we see straightforwardly that $\sin 2\theta_M \to 0; \quad \cos 2\theta_M \to -1$. This means, that $\theta_M \to \frac{\pi}{2}$. So we do not have oscillations in this case:

$$P_{matter}(\nu_e \to \nu_\mu; t)\bigg|_{N_e \gg \frac{\Delta m^2}{2\sqrt{2}|\vec{p}|G_F}} \to 0 \qquad (4.57)$$

Actually one sees directly that $H_{matter}$ in this case is diagonal. The very interesting thing here is one that the electron neutrino weak interaction eigenstate in matter coincides with the Hamiltonian eigenstate $\left|\nu_2^M\right\rangle$ now:

$$\left|\nu_e\right\rangle = \cos\theta_M \left|\nu_1^M\right\rangle + \sin\theta_M \left|\nu_2^M\right\rangle \xrightarrow{\theta_M \to \frac{\pi}{2}} \left|\nu_2^M\right\rangle. \qquad (4.58)$$

## 4.4 Atmospheric neutrino oscillations

The flux of neutrinos produced in the atmosphere is mostly produced through the decay of pions:

$$\begin{aligned} \pi &\to \nu_\mu \mu \\ &\hookrightarrow \nu_\mu \nu_e e \end{aligned} \qquad (4.59)$$

Theoretically we expect $\nu_\mu$ (~ 66%) and $\nu_e$ (~33%). Large underground detectors, originally conceived to search for proton decay, are sensitive to these atmospheric neutrinos. However, since the early 1990's the observed flux of muon neutrinos appeared to be much smaller than expected, the ratio $R$ is about [1]:

$$R = \frac{\left(\frac{\nu_\mu}{\nu_e}\right)_{observed}}{\left(\frac{\nu_\mu}{\nu_e}\right)_{predicted}} \approx 0.6 \qquad (4.60)$$

The neutrino oscillations could be the explanation for this anomalous ratio. Especially in summer 1998 from the superKamiokande experiment we have strong evidence for neutrino oscillations. This is the zenith angle dependence for the multi-GeV muon neutrinos. For neutrinos with energies in the multi-GeV range, the neutrino fluxes are not affected by geomagnetic effects in an asymmetric fashion [1]. So one expects the up-down symmetric neutrino signal. As we can see in Figure (4.2), in contrast to the 256 down-going muon neutrino events there are only 139 up-going events [1].

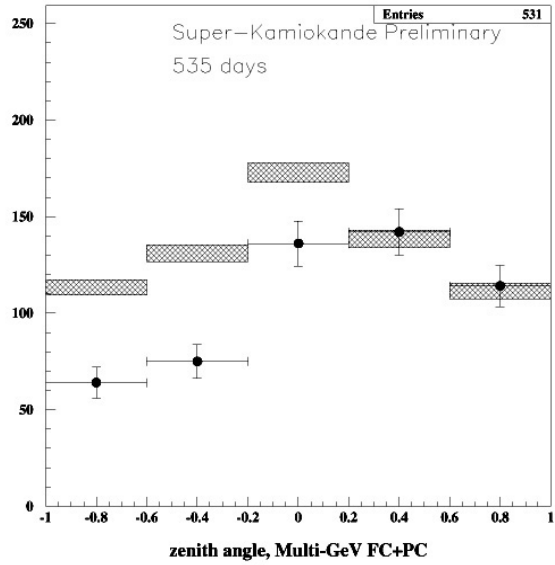

Figure (4.2) SuperKamiokande results on multi-GeV muon events.

This is clearly up-down asymmetric and the observed asymmetry is given by [1]:

$$\left(\frac{U-D}{U+D}\right)^{Multi-GeV}_{\nu_\mu} = -0.296 \pm 0.048 \pm 0.010. \qquad (4.61)$$

One notes that the *U* denotes up and the *D* means Down. For the electron neutrinos is the ratio like the following:

$$\left(\frac{U-D}{U+D}\right)^{Multi-GeV}_{\nu_e} = -0.036 \pm 0.067 \pm 0.020. \qquad (4.62)$$

This result is consistent with zero.

The SuperKamiokande collaboration interprets these results as evidence for $\nu_\mu \to \nu_x$ oscillations, $\nu_x$ means some other neutrino species. From the above fact and analysis, we can suppose that the electron neutrino events will not changed with the change of the ratio $L/E_\nu$ and in opposite to the electron neutrinos, the muon neutrino events must be changed in this case. The figure (4.3) demonstrated this situation clearly. One sees that the data from the SuperKamiokande experiment dramatically correspond with the computed data generated with the Monte Carlo method. One recalls the equation (4.25), and will suggest immediately that $\Delta m^2 \sim 10^{-3}$ eV$^2$. Actually after a more detailed analysis one shows that the best-fit data are $\sin^2 2\theta = 1$ and $\Delta m^2 = 2.2 \times 10^{-3}$ eV$^2$ [1].

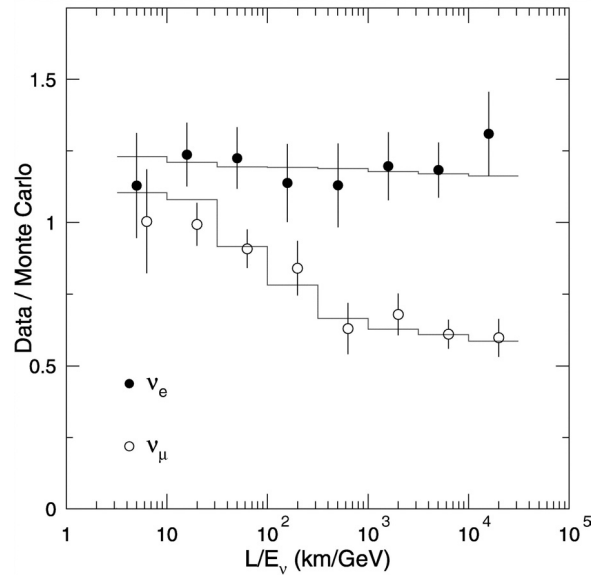

Figure (4.3)   Plot of neutrino signals in the SuperKamiokande experiment as a function of L/E$_\nu$

The most likely oscillations in the SuperKamiokande experiment should be the muon neutrinos to taon neutrinos, since the region in the $\Delta m^2$ - $\sin^2 2\theta$ plane, where muon neutrino to electron neutrino oscillations will be favored by the SuperKamiokande experiment, seems totally excluded by the null results of the CH00Z reactor experiment [1]. Furthermore, when there are oscillations between muon neutrinos and electron neutrinos, we should expect another asymmetry ratio $(U-D)/(U+D)$ [1].

Figure (4.4) summarizes all information about the evidence of the $\nu_\mu \rightarrow \nu_x$ oscillations. In general the experiments are consistent with each other.

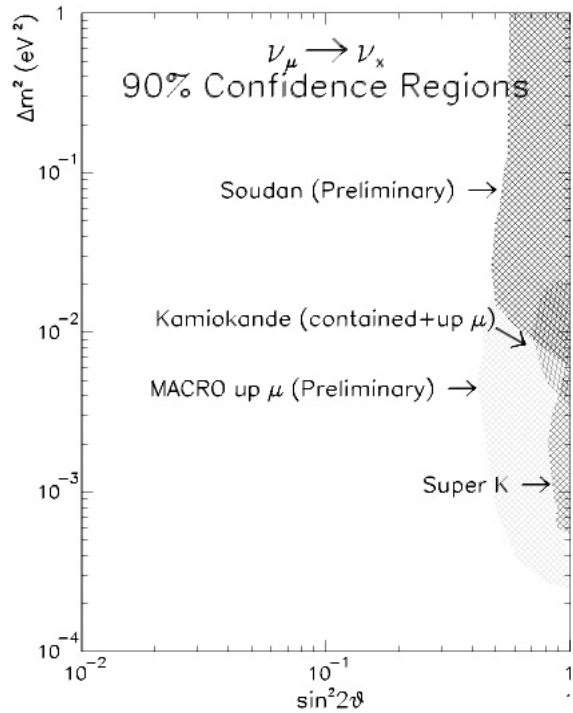

Figure (4.4) Evidence for atmospheric neutrino oscillations from all experiments, from [1]

## 5 Neutrinos in astrophysics and cosmology

In this chapter we concentrate on the neutrinos in astrophysics and cosmology. We have more data for the evidence of the neutrino oscillations and also some constraints for the neutrino mass.

### 5.1 The standard solar model and the solar neutrinos

First we introduce the Standard Solar Model (SSM) briefly. The SSM is based on the following observed parameters [9]:

| | |
|---|---|
| Surface Luminosity | $L^\odot = 3.86(1 \pm 0.005) \times 10^{33}$ erg/sec |
| Surface Temperature | $T_s^\odot = 5.78 \times 10^3$ K |
| Solar Mass | $M^\odot = 1.99 \times 10^{33}$ g |
| Solar Radius | $R^\odot = 6.96 \times 10^5$ Km |

Table (5.1) Observed Sun parameters

It describes the Sun as having been shining steadily and quietly for 4.6 billion years. It also assumes that the sun is spherically symmetric, in hydrostatic and thermal equilibrium, and is described by the ideal gas equation of state, with an initial abundance of elements similar to its primordial composition.

As everybody knows, the sun produces energy by nuclear reaction. According to the SSM, the pp chain is responsible for 98.5% of the energy production, whereas the CNO cycle produces only 1.5% of the total energy output. This ratio is very sensitive to the core temperature. For example, if the core temperature ($T_c \sim 1.56 \times 10^7$) of the sun was higher than $1.8 \times 10^7$ K, the CNO cycle would be the dominant energy-producing process.

The sun consists of approximately 70.5% protons (in mass), 27.5% $^4$He and 2% "heavy elements" which are elements heavier than $^4$He. The entire sun is roughly divided into three zones: (i) Solar Core (where $r \leq 0.3$ R$^\odot$), (ii) Radiation Zone (where $0.3$ R$^\odot \leq r \leq 0.71$ R$^\odot$) and (iii) Convection Zone (where $0.71$ R$^\odot \leq r \leq$ R$^\odot$).

After this short introduction about the SSM, let us consider the neutrinos. The possibility of observing solar neutrinos began to be discussed seriously following the 1958 experimental discovery by Holmgren and Johnston that the cross section for production of the isotope $^7$Be by the fusion reaction $^3$He + $^4$He → $^7$Be + γ was more than a thousand times larger than was previously believed. This result led Willy Fowler and Al Cameron to suggest that $^8$B might be produced in the sun in sufficient quantities by the reaction $^7$Be + p → $^8$B + γ. This $^8$B will produce an observable flux of high-energy neutrinos from the beta decay reaction $^8$B → $^8$Be + e$^+$ $\nu_e$.

The study of the solar neutrino flux was started in the early 1970's by Ray Davis and his group [1]. At present there are five different experiments, which give information on solar neutrinos (Gallex, Homestake, Kamiokande, SuperKamiokande, and SAGE). All these five experiments have some bearing on the issue of neutrino oscillations. Roughly said, all these experiments see only about 50% of the expected rate. This is demonstrated in Figure (5.1):

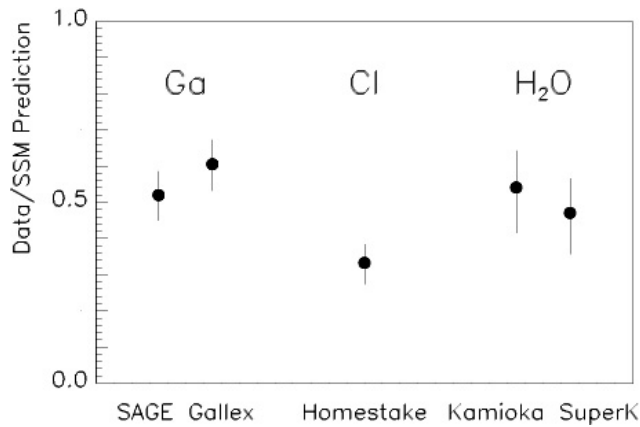

Figure (5.1)  Rates seen by the different solar neutrino experiments, compared to the expectations of the standard solar Model [1]

There are two distinct neutrino oscillation solutions to the solar neutrino problem. Because roughly all experiments are reduced by a factor of two from the expectations, it's possible to fit the data by using vacuum neutrino oscillations $\nu_e \to \nu_X$. With the earth-sun distance $L \sim 10^{11}$ one has typically $\Delta m^2 \sim 10^{-11}$ eV$^2$. This is demonstrated in Figure (5.2) as "just so".

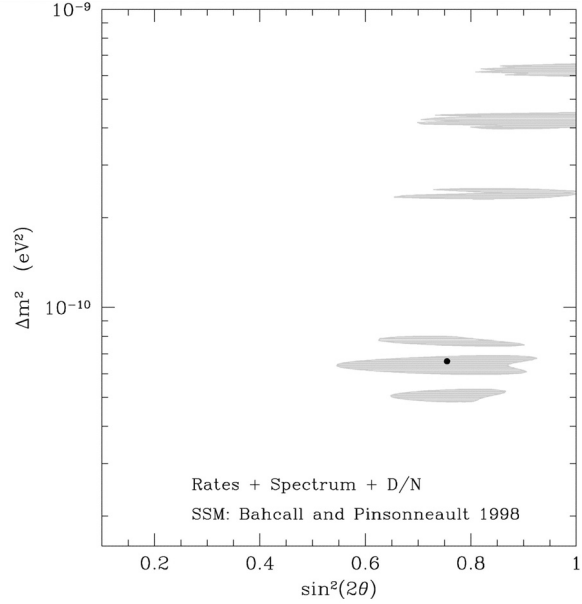

Figure (5.2) "Just so" solar neutrino fit [1]

Another is the MSW effect solution. In the sun the electron density can be characterized by an exponential profile function [1]:

$$N_e(r) = N_e(0) e^{-\frac{10r}{R_0}} \qquad (5.1)$$

The central density $N_e(0) \approx 10^{26}$ cm$^{-3}$ $\approx 10^{12}$ (eV)$^3$ is rather high and for appropriate values of $\Delta m^2$ and $\sin^2 2\theta$ can exceed the critical MSW density. For example: $\Delta m^2 \approx 10^{-5}$ eV$^2$ and p ~ 3 MeV and from the equation (4.52) we get the critical density about $10^{11}$ (eV)$^3$.

We recall the equation (4.47) with $N_e = N_e(t)$, so now is the equation (5.2) time dependent.

$$H_{matter}(t) = \frac{\Delta m^2}{4|\vec{p}|} \sin 2\theta \cdot \sigma_1 - \left( \frac{\Delta m^2}{4|\vec{p}|} \cos 2\theta - \frac{G_F N_e(t)}{\sqrt{2}} \right) \cdot \sigma_3 \quad (5.2)$$

One can diagonalize this Hamiltonian, the resulting energies and mixing angles are as following [1]:

$$E_{1,2}^M = \pm \sqrt{\left( \frac{\Delta m^2}{4|\vec{p}|} \cos 2\theta - \frac{G_F N_e(t)}{\sqrt{2}} \right)^2 + \left( \frac{\Delta m^2}{4|\vec{p}|} \sin 2\theta \right)^2} \qquad (5.3)$$

$$\tan 2\theta_M(t) = \frac{\frac{\Delta m^2}{4|\vec{p}|}\sin 2\theta}{\frac{\Delta m^2}{4|\vec{p}|}\cos 2\theta - \frac{G_F N_e(t)}{\sqrt{2}}} \quad (5.4)$$

We assume at the solar core $N_e(0) \gg N_e^{crit}$, according to the equation (4.58) $|\nu_e\rangle \cong |\nu_2^M(0)\rangle$. Furthermore we assume there are no transitions in the sun, the so-called adiabatic condition. Consequently

$$P_{solar}^{adiabatic}(\nu_e \to \nu_e; L) = \sin^2\theta \quad (5.5)$$

A more careful analysis shows that there are two MSW solutions, one adiabatic and one non-adiabatic, [1], both having $\Delta m^2 \sim 10^{-5}$ eV$^2$. The adiabatic solution has large mixing angles $\sin^2 2\theta \sim 1$ and for the non-adiabatic solution has $\sin^2 2\theta \sim 5\times 10^{-3}$.

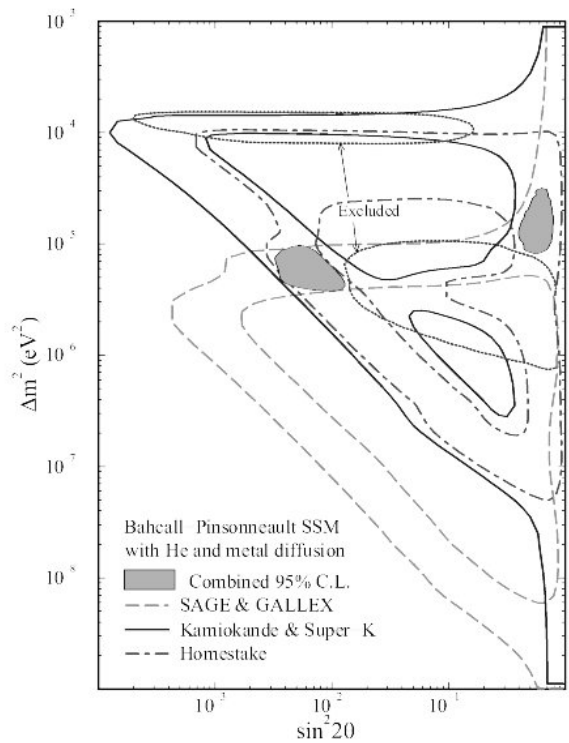

Figure (5.3) Regions in the $\Delta m^2$ – $\sin^2 2\theta$ plane favored by the MSW explanations of the solar neutrino data [1]

Figue (5.4) shows a more recent MSW analysis of the solar neutrino data including the latest Super-Kamiokande results [18]. While one can have both the small and large angle solutions to the rates and energy spectrum, only the small angle solution survives after including the zenith angle distribution.

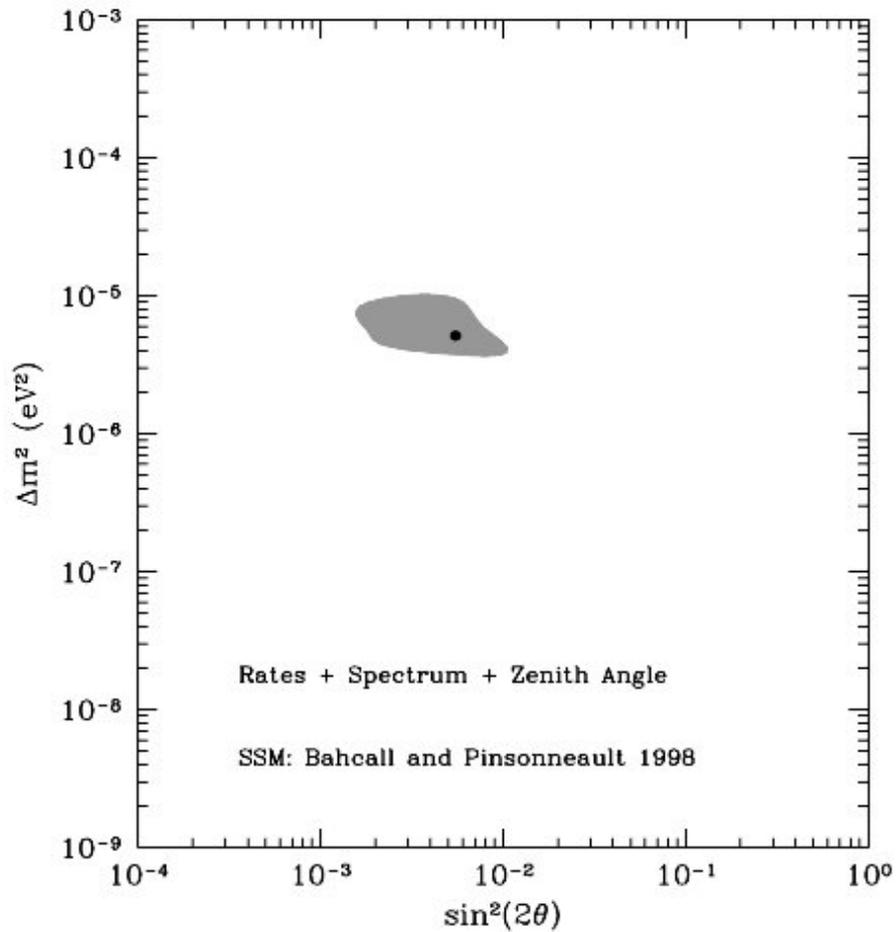

Figure (5.4) The MSW solution to the combined solar neutrino data on suppression rates along with the Super-Kamiokande energy spectrum and zenith angle distribution. The contour is drawn at 99% CL [18]

### 5.2 Cosmological constraints

We may recall from chapter 3, where we have discussed the see-saw mechanism, that the neutrino masses may split to one very light and one very heavy neutrino. This is actually consistent with the cosmological constraint we will discuss. We define

$$\Omega_\nu = \frac{\rho_\nu}{\rho_c} \tag{5.6}$$

where

$$\rho_\nu = n_\nu \sum_i m_{\nu_i} \quad \text{and} \quad \rho_c = \frac{3H_0^2}{8\pi G} \tag{5.7}$$

The $H_0$ is written in [7] as follows:

$$H_0 = 100 h \frac{km}{\sec \cdot Mpc} \tag{5.8}$$

Because of the lack of precise knowledge of $H_0$ the "lower-case h" cannot be indicated more accurately than

$$0.4 \leq h \leq 1.0 \tag{5.9}$$

(Historically Hubble's initial determination was $H_0 = 550$ km sec$^{-1}$ Mpc$^{-1}$.) With some recent paper, one has:

$$h = 0.65 \pm 0.1 \tag{5.10}$$

So the $\Omega_\nu$ must be less than one, in other words the $\rho_\nu$ must be less than the $\rho_c$. These considerations apply to the limits for the neutrino masses. To derive this limit, we have to find out the number density of neutrinos in the present universe. This depends on whether the neutrinos are relativistic or non-relativistic at the time of decoupling, so called hot relics or cold relics. We know, when $\Gamma$, the interaction rate for the reactions that keep the species in thermal equilibrium, is less than H, then they will decouple. We now roughly calculate the decoupling temperature for neutrinos as follows, since

$$\Gamma = n_\nu \langle \sigma v \rangle \sim G_F^2 T^5 \tag{5.11}$$

with the Fermi constant $G_F \sim 10^{-5}$ GeV$^{-2}$, decoupling occurs at a temperature $T_D$, when the condition $\Gamma \sim H$ is fulfilled, i.e. when $H \sim T^2/M_P$. This gives the following result:

$$T_D \approx \left(\frac{1}{G_F^2 M_P}\right)^{1/3} \sim 1 MeV \tag{5.12}$$

So we can also define that neutrinos are hot relics if they have a mass much less than $T_D$. When the mass is much greater than $T_D$, the neutrinos become non-relativistic while in equilibrium; then they are cold relics. For the hot relics the number density is similar as for photons, we find an upper limit for the mass for the light stable neutrinos. As for the cold relics, their number density falls exponentially with temperature as exp(-m/T). This case, when the neutrinos decouple, their number density is already so small that its contribution to mass density need not exceed the critical density. Therefore we find a lower limit for the heavy stable neutrinos. (However, in most gauge models for massive neutrinos, we expect the heavier neutrinos to be unstable. One has found, when the massive neutrinos were unstable and had a sufficiently short lifetime, then the limit can be avoided [8].) Figure (5.5) shows the cosmological constraints on stable neutrinos.

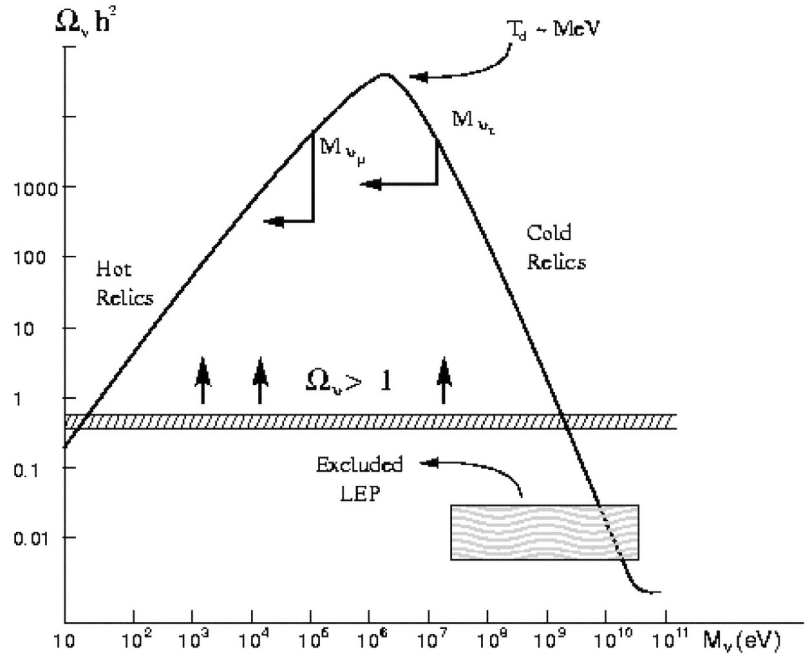

Figure (5.5)  Plot of $\Omega_\nu h^2$ as a function of the neutrino (sum) mass [1]

## 6   Summary for neutrino experiments

In this chapter we summarize some resent experimental results. Actually, besides neutrino oscillation phenomena, there is really no other direct evidence for neutrino masses. So certainly this summary discusses only the oscillation experiments.

In figure (6.1) we see the confusing situation. One sees a combination of the BNL 776 data and the Bugey reactor data. These only leave the region between 0.2 eV$^2$ < $\Delta m^2$ < 4 eV$^2$ as an "allowed" region for the LSND signal. However, this region is essentially excluded by the KARMEN2 data. This situation was discussed in some detail in the summary talk of Janet Conrad at the 1998 Vancouver International Conference on High Energy Physics [1].

So it is difficult to make strong statements at this stage. The best what one can say is that there are hints for muon neutrino to electron neutrino oscillations in the region 0.2 eV$^2$ < $\Delta m^2$ < 4 eV$^2$, with a small mixing angle $\sin^2 2\theta \sim 10^{-2}$.

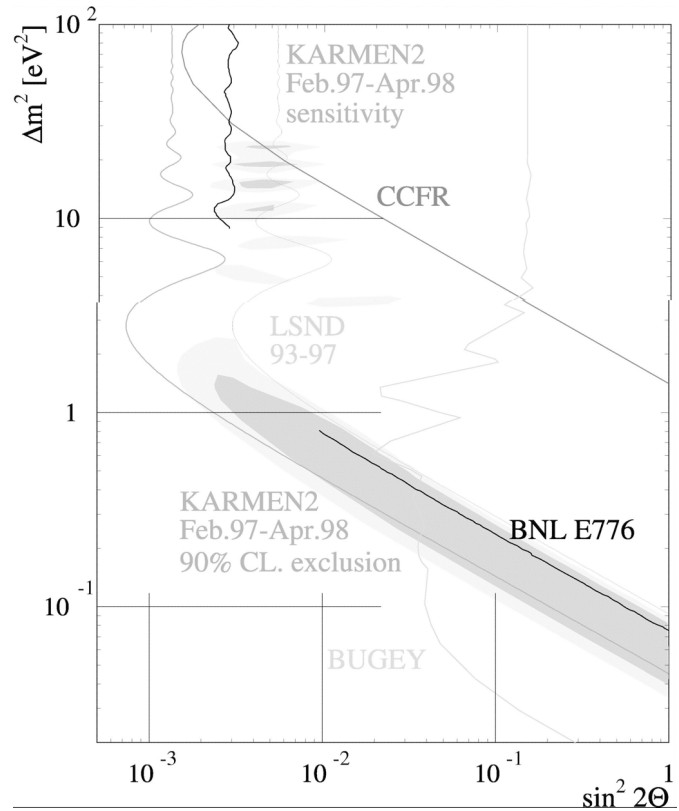

Figure (6.1) Summary of the experimental situation for $\bar{\nu}_\mu \to \bar{\nu}_e$ oscillation at tome of the Vancouver Conference. [1]

From [1] we have a collection of all the neutrino oscillation evidence, as well as the hints for oscillations, which we have at the moment. This is demonstrated in Figure (6.2), because of the discussion before, we do not recognize the LSND as a suggested oscillation region. So we have three regions suggested in the $\Delta m^2$ - $\sin^2 2\theta$ plane. The strongest evidence is that for atmospheric neutrino oscillations coming from the SuperKamiokande zenith angle data. Here the suggested parameters are $\Delta m^2 \sim 3\times 10^{-3}$ eV$^2$, $\sin^2 2\theta \sim 1$ with $\nu_\mu \to \nu_X$ ($\nu_X \neq \nu_e$). For solar neutrinos we have two solutions, one is the MSW $\nu_e \to \nu_X$ oscillation with the parameters $\Delta m^2 \sim 10^{-5}$ eV$^2$, $\sin^2 2\theta \sim 1$ (excluded from [18]) or $\sin^2 2\theta \sim 5\times 10^{-3}$, or the "just so" $\nu_e \to \nu_X$ oscillation with the parameters $\Delta m^2 \sim 10^{-11}$ eV$^2$, $\sin^2 2\theta \sim 1$.

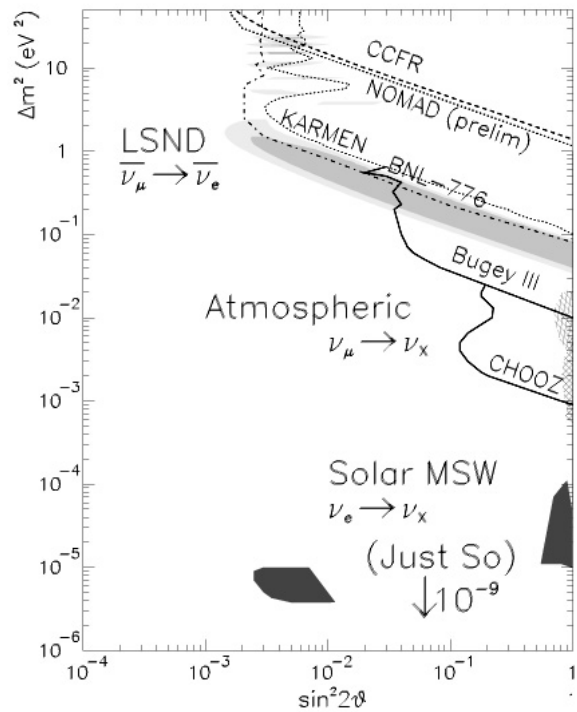

Figure (6.2) Summary of evidence of, and hints for, neutrino oscillations [1]

Until now we still do not have a definite answer to the question, whether neutrinos have mass. We hope that in near future we will have more evidence and will be able to answer this question.

**Acknowledgements**

I would like to thank my tutor Igor Tkachev for his guide and suggestions. I would like to express my thanks to my colleges Jan Kratochvil and Markus Engeli for the comments and criticism. I am very grateful to Craeg K. Strong of MIT for the English corrections.